\documentclass[aps,pre,reprint,superscriptaddress,doi=false,isbn=false,url=false,title=true,longbibliography,noeprint,footinbib]{revtex4-1}

\usepackage{graphicx}
\usepackage{amsmath}
\usepackage{latexsym}
\usepackage {mathrsfs}
\usepackage{float}
\usepackage{amssymb} 
\usepackage[utf8]{inputenc}
\usepackage{cmap} 
\usepackage[T1]{fontenc}
\usepackage{microtype}
\usepackage{amsmath,amssymb,mathtools}
\usepackage{txfonts}

\usepackage{mathrsfs}
\usepackage{graphicx,grffile,textcomp}
\graphicspath{.}
\usepackage{soul}
\usepackage{textcomp}
\usepackage{booktabs,tabularx,dcolumn}
\newcolumntype{d}[1]{D{.}{.}{#1}}
\usepackage{url,hyperref}
\usepackage[usenames,dvipsnames]{xcolor}
\hypersetup{colorlinks=true, linkcolor=BrickRed, urlcolor=blue!50!black, citecolor=blue!50!black}
\usepackage{textcomp}
\usepackage {mathrsfs}
\usepackage{xcolor}

\begin{document}
\title{Domain Growth and Aging in a Phase Separating Binary Fluid Confined Inside a Nanopore}
\author{Saikat Basu}\email[]{saikatjuphy0809@gmail.com}
\affiliation{ Department of Physics, Sonamukhi College, Bankura, West Bengal 722202, India}
\author{Suman Majumder}\email[]{suman.jdv@gmail.com}   
\affiliation{Amity Institute of Applied Sciences, Amity University Uttar Pradesh, Noida 201313,
India}
\author{Raja Paul}\email[]{ssprp@iacs.res.in}
\affiliation{Indian Association for the Cultivation of Science, Kolkata 700032, India}
\author{Subir K. Das}\email[]{das@jncasr.ac.in}
\affiliation{Theoretical Sciences Unit and School of Advanced Materials, Jawaharlal Nehru Centre for Advanced Scientific
Research, Jakkur P.O, Bangalore 560064, India}
\date{\today}

\begin{abstract}
Hydrodynamics is known to have strong effects on the kinetics of phase separation. There exist open questions on how such effects manifest in systems under confinement. Here, we have undertaken extensive studies of the kinetics of phase separation in a two-component fluid that is confined inside pores of cylindrical shape. Using a hydrodynamics-preserving thermostat, we carry out molecular dynamics simulations to obtain results for domain growth and aging for varying temperature and pore-width. We find that all systems freeze into a morphology where stripes of regions rich in one or the other component of the mixture coexist in a locked situation. Our analysis suggests that, irrespective of the temperature the growth of the average domain size, $\ell(t)$, prior to the freezing into stripped patterns, follows the power law $\ell(t)\sim t^{2/3}$, suggesting an inertial hydrodynamic growth, which typically is applicable for bulk fluids only in the asymptotic limit. Similarly, the aging dynamics, probed by the two-time order-parameter autocorrelation function, also exhibits a temperature-independent power-law scaling with an exponent $\lambda \simeq 2.55$, much smaller than what is observed for a bulk fluid.
\end{abstract}

\maketitle 

\section{Introduction}\label{intro} 
Understanding of structure and dynamics of fluids in confined systems is of significant importance in the context of nanoscience and nanotechnology ~\cite{evans1990liquids, gelb1999phase,thorsen2002microfluidic, brovchenko2008interfacial,zhang2017nanoconfined}. Fundamentally, the observed phenomena can be linked to surface-induced heterogeneous nucleation or wetting, which is often directly connected to thermodynamics and transport in various disordered systems and complex media \cite{havlin1987diffusion,paul2005condensation,schehr2005universal,henkel2006ageing,burioni2006aging,paul2007superaging,henkel2008superuniversality,Park2010,horikawa2011capillary,bear2013dynamics,corberi2013scaling}. Studies of such systems can have significant micro- and nano-fluidic applications, for example, in oil extraction from rocks.
\par
An interesting set-up is when a system under confinement is inside the multi-phase coexistence by virtue of the choice of parameters like temperature ($T$), pressure ($P$), density ($\rho$), or composition in a multicomponent mixture \cite{jones1991surface}. If the underlying state is fluid, the effects of hydrodynamics during the kinetics of phase separation in such systems are of paramount importance \cite{binder2010phase,bray2002theory}. Despite their practical relevance, such problems in these systems have been studied less extensively than in bulk situations. Most existing works have focused either on systems confined between parallel plates~\cite{Jones1991,monette1992,puri1997surface,Binder_rev,puri2001power,puri2005surface,das2006molecular,das2006spinodal,binder2010phase,jaiswal2012hydrodynamic} or on two-dimensional nanopores~\cite{paul2005condensation}. The consideration of curved boundaries, therefore, represents a step forward, reflecting situations frequently encountered in nature.
\par
For confinement between parallel plates, varying the plate separation allows one to switch between space dimensions $d=2$ and $d=3$. In contrast, for cylindrical confinement, the transition occurs between \cite{gelb1999phase,zhang1994phase,gelb1997kinetics,gelb1997liquid,basu2017phase,davis2023surface} $d=1$ and $d=3$. Cylindrical geometries with curved boundaries also have fundamental significance; for instance, the interfacial tension between phases can be controlled by the radius of curvature, which in turn can influence the dynamics, particularly when the radius is sufficiently small \cite{majumder2010,majumder2011diffusive}.
\par

To study the dynamics of phase separation in a binary (A+B) fluid mixture, a system is usually prepared above the demixing critical temperature $T_c$ and then quenched below the coexistence curve \cite{bray2002theory}. After the quench, the system becomes unstable to fluctuations and evolves toward a phase-separated state \cite{bray2002theory}. The evolution begins with the formation of domains of similar species, which subsequently grow over time. In a bulk system, the corresponding average domain size, $\ell$, increases with time ($t$) as \cite{bray2002theory}
\begin{equation}\label{power-law}
\ell(t) \sim t^\alpha. 
\end{equation}
The exponent $\alpha$ can depend on factors such as the transport mechanism, conservation and symmetry of the order parameter, and the spatial dimension \cite{bray2002theory}. In the absence of hydrodynamics, i.e., for purely diffusive transport with a conserved scalar order parameter ($\mathcal{O}$), as in phase separation of a binary mixture, the exponent takes the value  \cite{lifshitz1961kinetics} $\alpha = 1/3$. The presence of hydrodynamics modifies this scenario, and a single growth exponent may no longer describe the entire process. For instance, in $d=3$ one finds \cite{bray2002theory,lifshitz1961kinetics,siggia1979late,furukawa1985effect,furukawa1987turbulent}
\begin{eqnarray}\label{growthlaw}
  \alpha=\begin{cases}
    1/3, & \text{if $\ell(t) \ll \ell_{v}=(D\eta)^2$},\\
    1, & \text{if $\ell_{v} \ll \ell(t) \ll \ell_{in}$},\\
    2/3, & \text{if $ \ell(t) > \ell_{in}$}.
  \end{cases}
\end{eqnarray}
Here $\ell_{v}$ and $\ell_{in}$, having possible dependence on certain diffusivity $(D)$ and viscosity $(\eta)$, are lengths characterizing crossovers from diffusive coarsening to viscous hydrodynamic growth to inertial hydrodynamic relaxation ~\cite{gelb1999phase, Jones1991, puri1997surface, puri2001power, puri2005surface, binder2010phase}.


During such growth, the structure exhibits self-similarity, i.e., the systems at different times differ only by a change in the characteristic length scale $\ell$. Such self-similarity is reflected in the scaling behavior of the two-point equal-time correlation function \cite{bray2002theory},
\begin{eqnarray}\label{spacecor}
C(r,t) = \langle\mathcal{O}(\vec{r},t)\mathcal{O}(\vec{0},t)\rangle-\langle\mathcal{O}(\vec{r},t)\rangle\langle\mathcal{O}(\vec{0},t)\rangle,
\end{eqnarray}
as \cite{bray2002theory}
\begin{equation}\label{Scaling_Cr}
C(r,t)=\tilde{C}\left[ r/\ell(t)\right ],
\end{equation}
where $\tilde{C}$ is a function \cite{bray2002theory, puri2009kinetics} that is independent of time, with $r$ being the separation between the points. The corresponding form for the structure factor, the Fourier transform of the real-space function, is  \cite{bray2002theory}
\begin{equation}\label{Strfac}
S(k,t)=\ell^{-d}\tilde{S}(k\ell),
\end{equation}
where $k$ is the wave number and $\tilde{S}(k\ell)$ is another master function. Note that $\mathcal{O}(\vec{r},t)$,in Eq.\ \eqref{spacecor}, is a space ($\vec{r}$) and time dependent order parameter.
\par
Another notable feature of the evolution is the onset of physical aging, which appears as a deviation from time-translational invariance together with dynamical scaling \cite{Bouchaud_book,henkelbook,Zannetti_book}. This is typically examined through the two-time order-parameter autocorrelation function \cite{fisher1988}: 
\begin{eqnarray}\label{autocor}
C_{\rm{ag}}(t,t_{w}) = \langle\mathcal{O}(\vec{r},t)\mathcal{O}(\vec{r},t_{w})\rangle-\langle\mathcal{O}(\vec{r},t)\rangle\langle\mathcal{O}(\vec{r},t_{w})\rangle,
\end{eqnarray}
where $t$ and $t_{w}$ ($<t$), respectively, are the observation and waiting times. Above mentioned dynamical scaling can be realized when $C_{\rm{ag}}(t,t_{w})$ is plotted as a function of $t/t_w$ or $\ell / \ell_{w}$, $\ell_{\rm{w}}$ being the value of characteristic length at time $t_{\rm{w}}$. Typically, the form of the scaling is a power-law ~\cite{fisher1988,yeung1996,Liu1991, Lorenz2007, Midya2014, Midya2015, roy2019aging}, at least asymptotically $(\ell / \ell_{\rm{w}} \rightarrow \infty)$, i.e.,
\begin{eqnarray}\label{corscaling}
C_{\rm{ag}}(t,t_{w}) \sim x^{-\lambda}; ~ x=\ell(t)/\ell(t_{w}).
\end{eqnarray}
Here $\lambda$ is the autocorrelation exponent. Yeung, Rao and Desai (YRD) obtained a bound for this quantity ~\cite{yeung1996}, viz,
\begin{equation}\label{YRD-bound}
 \lambda \geq \frac{\beta+d}{2},
\end{equation}
where $\beta$ is an exponent \cite{yeung1988scaling} concerning the power-law enhancement of $S(k,t_w)$ in the small $k$ limit, i.e.,
\begin{equation}\label{Strfacbeta}
S(k \rightarrow 0,t_{\rm{w}}) \sim k^{\rm{\beta}}.
\end{equation}
The bound \label{YRD-bound} is a general version of the one provided by Fisher and Huse \cite{fisher1988}.
\par
For bulk fluids, it is well established that hydrodynamics plays an important role not only in equilibrium dynamics ~\cite{sengers2000equations,onuki2002phase,bhattacharjee2010sound,das2006critical,roy2016structure} but also in the nonequilibrium kinetics of phase separation~\cite{Ahmad2010,majumder2011VL,das2012finite}, including aging dynamics \cite{roy2019aging}. Motivated by this, we focus here on the aging behavior associated with the kinetics of a phase-separating binary (e.g., comprising components A and B) fluid confined within a cylindrical pore. The investigation is performed using molecular dynamics (MD) simulations \cite{allen1987tildesley} with a temperature controller that preserves hydrodynamics.

The remainder of the article is structured as follows. Section~\ref{model} describes the model and simulation methods. Section~\ref{results} presents the results, and Section~\ref{conclusion} concludes with a summary and a brief discussion of future perspectives.

\section{Model and Methods}\label{model}
In this work, we set up a computational framework based on the protocol outlined in our previous study~\cite{basu2017phase}. The binary mixture~\cite{das2006critical} consists of spherical particles A and B, each with diameter $\sigma \,(=1)$ and mass $m \,(=1)$. Two particles $i$ and $j$, separated by a distance $r_{ij}$, interact via \cite{das2006critical}
\begin{eqnarray}\label{pot}
U(r_{ij}) = V(r_{ij}) - V(r_{c}) - (r_{ij} - r_{c}) \left(\frac{dV(r_{ij})}{dr_{ij}} \right) _{r_{ij}=r_{c}}. 
\end{eqnarray}
Here $V(r_{ij})$ is the full Lennard-Jones (LJ) interaction having the form \cite{allen1987tildesley}
\begin{eqnarray}\label{LJ_pot}
V(r_{ij})=4\epsilon_{\gamma,\delta}\left[\left(\frac{\sigma}{r_{ij}}\right)^{12} - 
\left(\frac{\sigma}{r_{ij}}\right)^{6}\right];~\gamma, 
\delta \in [{\rm A, B}].
\end{eqnarray}
To drive phase separation between A and B particles, the interaction strengths are chosen \cite{das2006critical} such that $\epsilon_{\rm{AA}}=\epsilon_{\rm{BB}}=2\epsilon_{\rm{AB}}=\epsilon$ with, for simplicity, $\epsilon=1$. This parameter choice introduces an Ising-like symmetry. For computational efficiency, the potential in Eq.~\eqref{pot} is truncated and shifted \cite{das2006critical} to zero at $r=r_c \,(=2.5\sigma)$. The additional term ensures continuity of force as well as potential, preventing energy jumps during MD simulations~\cite{allen1987tildesley,das2006critical}. At an overall particle density $\rho=1$, the model yields a well-established bulk critical temperature~\cite{das2006critical,roy2016structure} of $T_c \simeq 1.42\,\epsilon/k_B$. Throughout this study, the Boltzmann constant, $k_{\rm B}$, is set to unity.
\begin{figure}[t!]
\centering
\includegraphics*[width=0.45\textwidth]{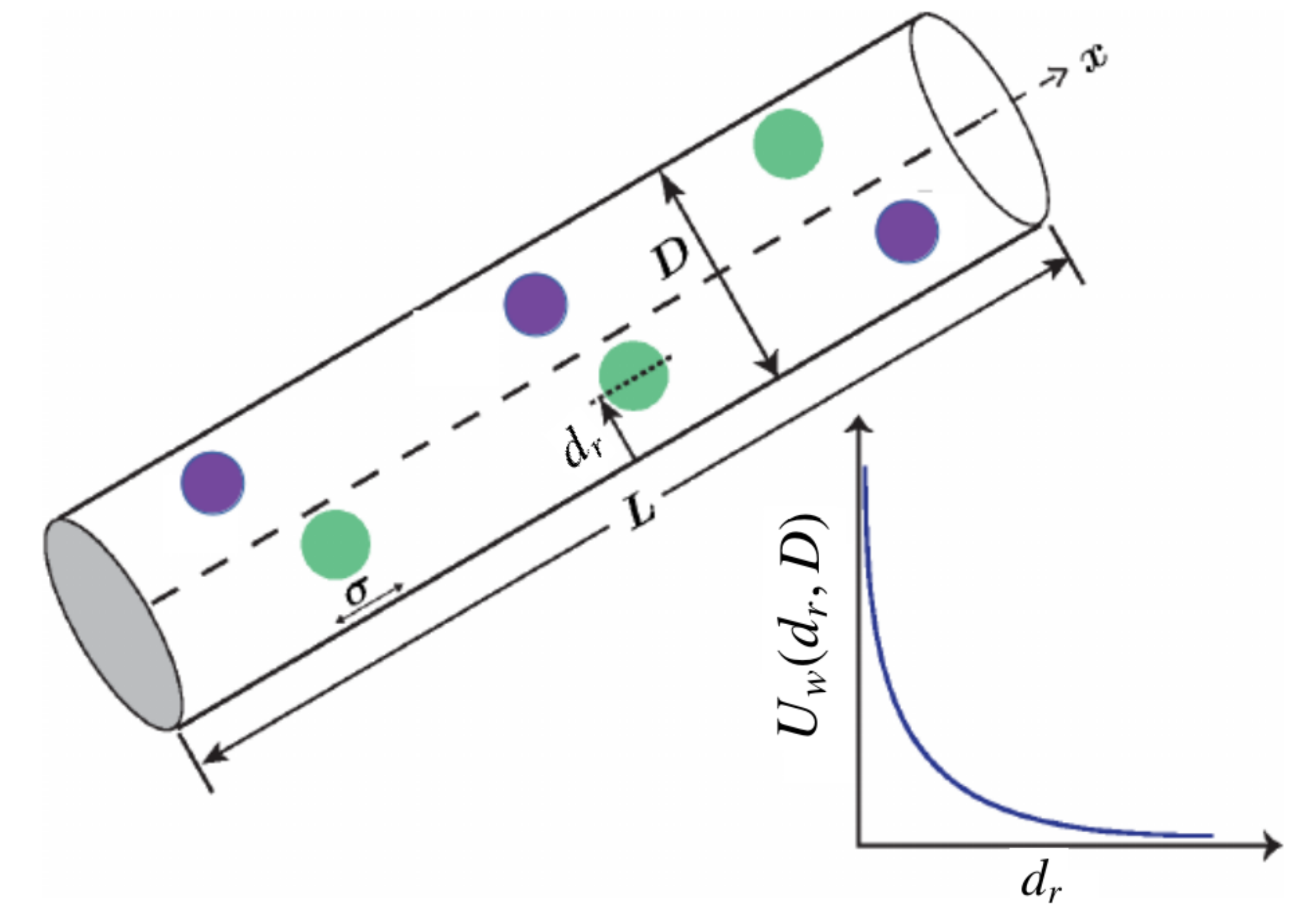}
\caption{Sketch of a model cylinder of length $L$ and diameter $D$. The wall consists of Lennard-Jones particles, while the interior contains A and B particles shown in different colors. The schematic plot below shows the interaction function between the cylindrical surface and a fluid particle separated by a distance $d_r$.}
\label{fig1}
\end{figure}
\par
We confine the binary mixture inside a model cylindrical nanopore \cite{basu2017phase}, illustrated schematically in Fig.~\ref{fig1}. The cylinder has a diameter $D$ that is much smaller than its length $L$, with all dimensions measured in units of $\sigma$. Along the axial ($x$) direction, periodic boundary conditions are imposed. The wall of the cylinder is represented by LJ particles but treated as structureless, since the potentials are integrated out. Consequently, a fluid particle located at a radial distance $d_r$ from the axis interacts with the wall via \cite{peterson1986fluid,kanda2007freezing}
\begin{eqnarray}\label{LJ_wall}
U_w(d_r,D)= \pi \rho_s \epsilon_w \left[ \frac{7}{32} \sigma^{12} K_9(d_r,D)- \sigma^{6} K_3(d_r,D) \right].
\end{eqnarray}
A schematic representation of the \textit{wall–particle} interaction potential, $U_w(d_r,D)$, as a function of the radial distance $d_r$, is shown in Fig.~\ref{fig1}. The wall–fluid interaction strength, $\epsilon_w (=0.1\epsilon)$, is neutral, meaning it does not preferentially interact with either component. In Eq.~\eqref{LJ_wall}, $\rho_s (=1)$ represents the density of interaction sites on the inner surface of the cylinder, and \cite{peterson1986fluid,kanda2007freezing}
\begin{eqnarray}\label{Kn}
\nonumber
&& K_n(d_r,D)=\left( \frac{2}{D} \right)^{n}\int_{0}^{\pi}d\theta\times\\ 
&& \left[ -\frac{2d_r}{D} \cos \theta +
\left( 1-4\left(\frac{d_r}{D}\right)^2 \sin^2 \theta \right)^{1/2} \right] ^{-n}.
\end{eqnarray}
The integral in Eq.~\eqref{Kn} cannot be solved analytically and is therefore evaluated with the aid of a numerically generated look-up table.

\par
Disordered configurations with equal proportions of A and B particles ($50:50$), prepared at high temperatures for an overall density $\rho = 0.8$, are quenched to $T = 1.1$ and lower. Since for bulk with $\rho = 1$, the critical temperature is ~\cite{das2012finite} $T_c \simeq 1.42$, this quench temperature is expected to be below $T_c$ also for $\rho = 0.8$. The simulations are performed in the canonical ensemble using molecular dynamics. To retain hydrodynamic effects~\cite{basu2017phase}, the Nos\'{e}-Hoover thermostat~\cite{frenkel2001} is employed for temperature control. The dynamical equations are solved with the velocity-Verlet algorithm~\cite{allen1987tildesley,frenkel2001}, using a time step $\Delta t = 0.01 \tau$, where $\tau = (m\sigma^2/\epsilon)^{1/2}$ is the Lennard-Jones time unit. Temperature is expressed in units of $\epsilon/k_B$, with $m, \sigma, \epsilon$, and $k_B$ all set to unity. For computing correlation functions and determining the characteristic length $\ell$, an order parameter is introduced in the next section. All numerical data are averaged over at least $100$ independent initial realizations.

\section{Results}\label{results}
This section is divided into two parts. In the first part, we present results on domain growth, expanding upon our earlier findings in Ref.~\cite{basu2017phase}. In the second part, we focus on aging phenomena, where we provide new results on this fundamental aspect.

\subsection{Growth Dynamics}\label{growth}
\begin{figure}[t!]
\centering
\includegraphics*[width=0.45\textwidth]{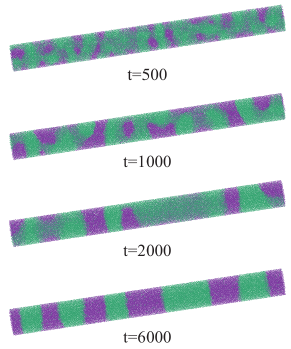}
\caption{\label{fig2} Evolution snapshots obtained after quenching a system, with $D=20$ and $L=200$,  from high temperature disordered phase to $T = 0.8$. Different species are marked in different colors. We have kept the cylinder length fixed throughout this study.}
\end{figure}

In contrast to bulk fluids, where a systen may reach a fully segregated equilibrium state, binary fluids confined within cylindrical pores often remain trapped in stripe-like arrangements of alternating A- and B-rich regions, as shown by several simulation studies \cite{gelb1997kinetics,gelb1999phase,basu2017phase}. This behavior is illustrated in Fig.~\ref{fig2}, where snapshots are presented for a system with $D=20$ and $L=200$ at $T=0.8$. At early times, small isotropic domains appear, initiating the segregation process. Growth at this stage proceeds through the usual diffusive mechanism. With time, these domains accumulate material and form stripes aligned periodically along the cylinder axis. Once the stripes attain a certain average width, it remains constant during the subsequent evolution. Such observations of ours are consistent with earlier findings on phase separation in confined pore geometries~\cite{zhang1994phase,gelb1997kinetics,gelb1997liquid,basu2017phase}, including Ising systems under similar confinement~\cite{Binder_rev,albano1992}, where the resulting configuration is often referred to as a plug-like phase \cite{monette1992}.
\par
It should be noted that, since the wall–particle interaction is neutral, wetting of the wall by either species is not expected. As a result, domains of both species extend across the cylinder width. At this temperature, thermal fluctuations are insufficient to form bridges between successive domains along the cylinder axis. This absence of bridging between alternating stripes, due to the weak interaction between adjacent stripe boundaries beyond a characteristic separation, is believed to contribute to the arrest of domain growth~\cite{basu2017phase}.

\begin{figure}[t!]
\centering
\includegraphics*[width=0.45\textwidth]{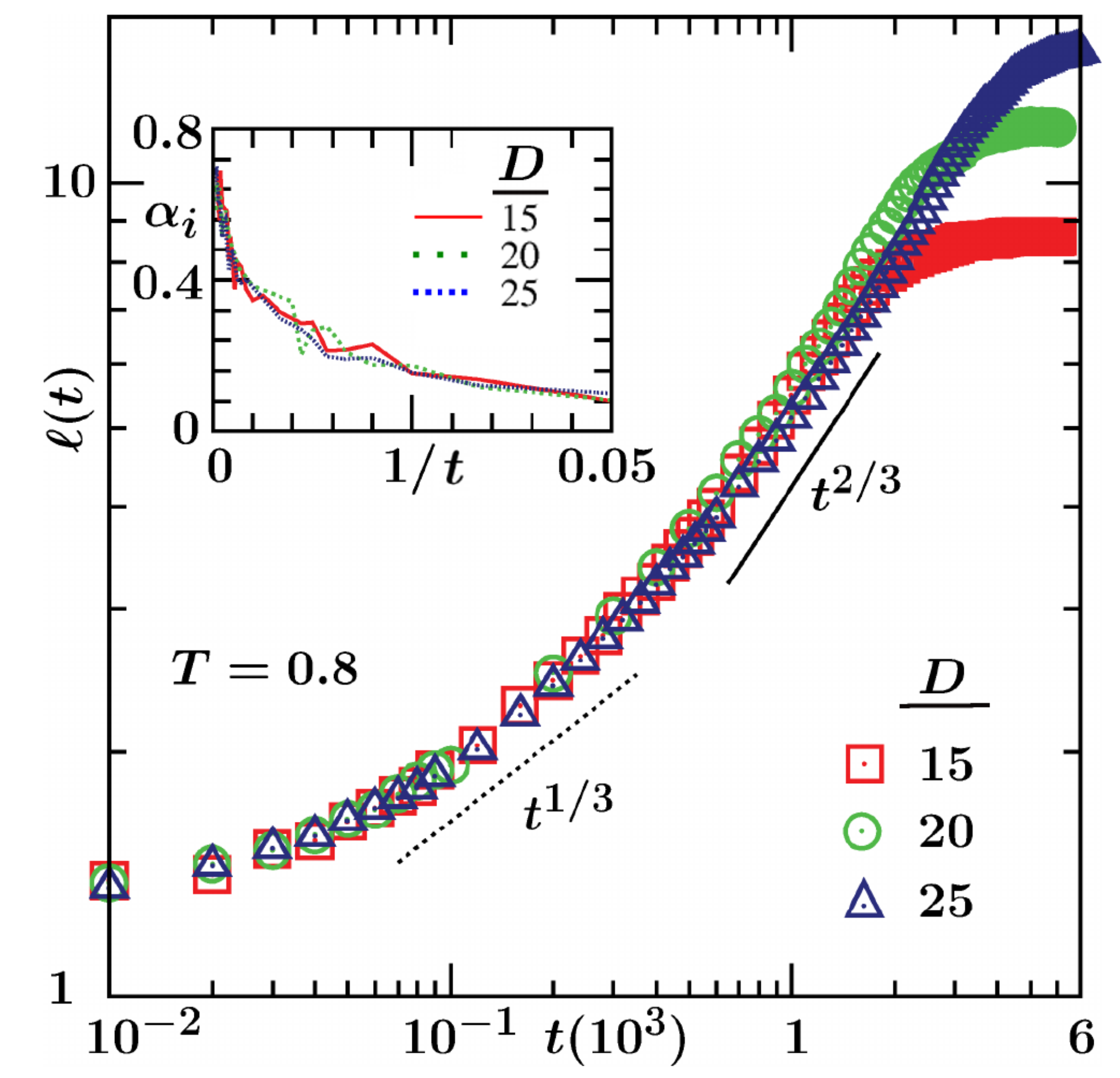}
\caption{\label{fig3} Plots of $\ell(t)$ (at $T=0.8$), as a function of time, on a log-log scale, where different colors and symbols are used to represent results for different diameter sizes, with $L$ fixed at $200$. The dotted line corresponds to diffusive ($\alpha=1/3$) and the solid line corresponds to inertial hydrodynamic ($\alpha=2/3$) growth laws. Inset: instantaneous exponent $\alpha_{i}$, corresponding to the data in the main frame, are plotted versus $1/t$.}
\end{figure}

\par 
Given the feature of axial, and therefore anisotropic, growth of domains, it is meaningful to quantify growth along the cylinder axis~\cite{zhang1994phase,gelb1997kinetics,gelb1997liquid,basu2017phase}. The domain length is calculated from the decay of the two-point equal-time correlation function~\cite{puri2009kinetics,bray2002theory}  
\begin{eqnarray}\label{CorrEq}
C(x,t) = \langle \mathcal{O}(0,t) \mathcal{O}(x,t) \rangle - \langle \mathcal{O}(0,t) \rangle \langle \mathcal{O}(x,t) \rangle,
\end{eqnarray}
where $\mathcal{O}(x,t)$ is the one-dimensional order parameter defined as  \cite{das2006spinodal,binder2010phase}
\begin{eqnarray}\label{Ordpar}
\mathcal{O}(x,t) = \frac{\rho_{\rm A}(x,t) - \rho_{\rm B}(x,t)}{\rho}.
\end{eqnarray}
Here, the angular brackets denote statistical averaging~\cite{basu2017phase}. For systems with a conserved order parameter, $C(x,t)$ typically exhibits damped oscillations around zero \cite{das2006spinodal,binder2010phase}. In the following, we use the distance at which $C(x,t)$ first crosses zero as a measure of the characteristic domain length, $\ell(t)$. In Eq.~\eqref{Ordpar}, $\rho_{\rm A}$ and $\rho_{\rm B}$ represent the local densities of species A and B, respectively.

To quantify the growth of stripes, Fig.~\ref{fig3} shows $\ell(t)$ as a function of $t$, for three pore diameters, viz., $D = 15, 20$, and $25$, plotted on a double-logarithmic scale, a standard approach when power-law behavior is anticipated. At early times, $\ell(t)$ follows the expected diffusive growth (dotted line). At later times, it crosses over to faster growth, appearing consistent with an exponent $\alpha = 2/3$ (solid line), for a power-law. The eventual saturation of $\ell(t)$ at $\ell \simeq D$ reflects the usual finite-size effect~\cite{Ahmad2010,majumder2011VL}. We note that the crossover between the two growth regimes is expected to occur at later times for higher temperatures, as thermal fluctuations enhance diffusivity. To better confirm the late-time growth, we estimate the instantaneous exponent~\cite{huse1986corrections,majumder2010,majumder2011diffusive}  
\begin{eqnarray}\label{InstExp}
\alpha_i = \frac{d \ln \ell(t)}{d \ln t}.
\end{eqnarray}  
Extrapolation of $\alpha_i$ in the limit $\ell \rightarrow \infty$ ($1/t \rightarrow 0$) provides the asymptotic value of $\alpha$. The inset of Fig.~\ref{fig3} illustrates this, confirming that $\alpha \simeq 2/3$, consistent with inertial hydrodynamic growth expected in the late stages of kinetics~\cite{bray2002theory} for bulk systems.
\begin{figure}[t!]
\centering
\includegraphics*[width=0.45 \textwidth]{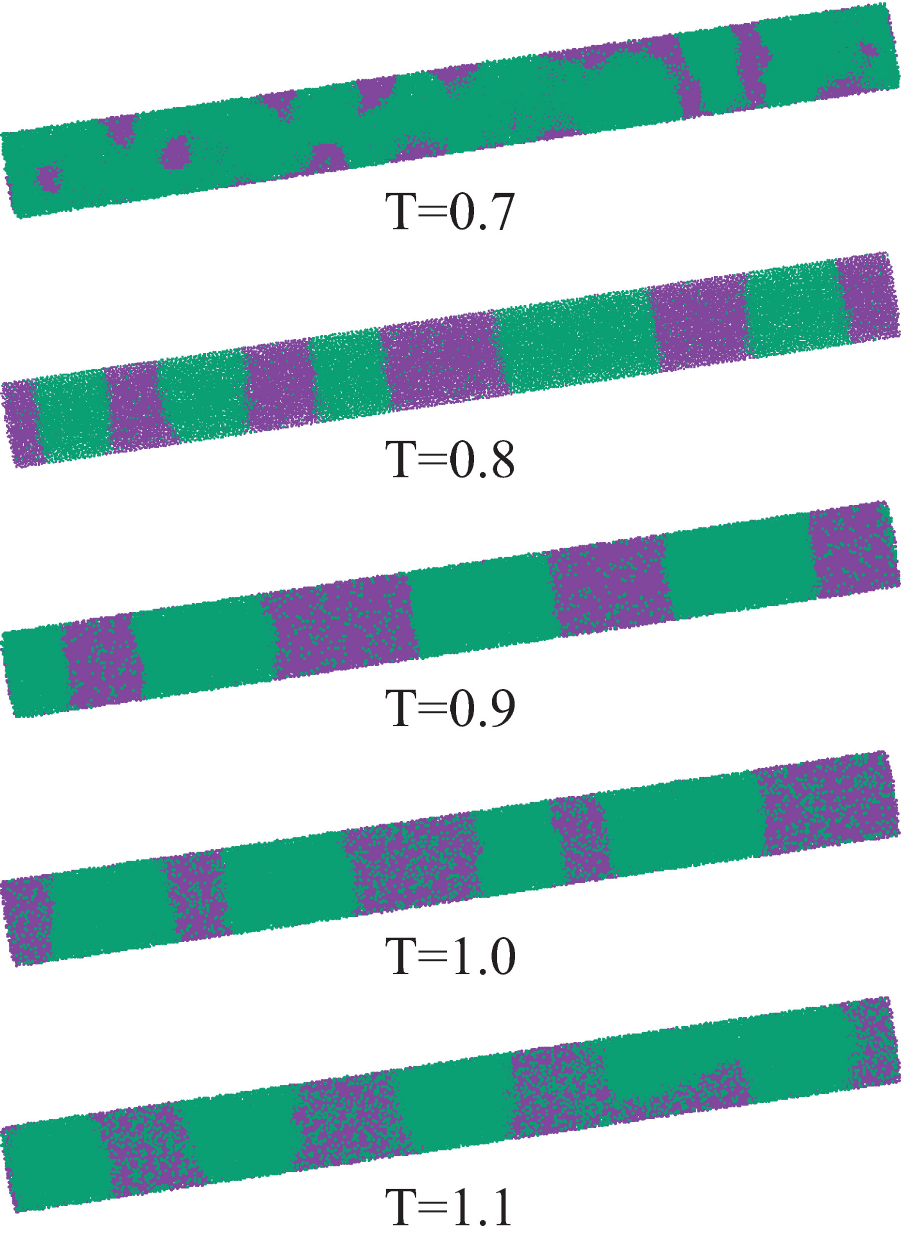}
	\caption{\label{fig4} Snapshots of the configurations obtained at $t=6000$ after quenching binary mixtures, with $D=20$ and $L=200$,  from high temperature phase to different final temperatures $T$, as indicated in the figure. Color coding is same as in Fig. \ref{fig2}.}
\end{figure}

All the results presented so far correspond to the final temperature $T = 0.8$. We now turn to examine the effect of changing the final temperature on the growth kinetics. In Fig.~\ref{fig4}, snapshots at a fixed time $t = 6000$ are shown for different final temperatures. Unless stated otherwise, all results regarding temperature dependence are for $D = 20$ and $L = 200$. The main mechanisms of domain coarsening remain similar to those discussed earlier across all final temperatures. At very early times, growth is driven by single-particle diffusion, while at later times, hydrodynamic effects increasingly contribute to the growth rate. For the lowest temperature considered, $T = 0.7$, domains do not span the full cylinder diameter, and domain growth is arrested in an intermediate metastable state. At higher temperatures, $T = 1.0$ and $1.1$, even at late times, the domains are not fully pure. Unlike the low-temperature case ($T = 0.8$), where domains are well-formed with sharp boundaries, the interfaces at $T \geq 1.0$ fluctuate. Nevertheless, these thermal fluctuations are insufficient to bridge the stripes and form a fully phase-separated morphology as in bulk systems. In conclusion, despite thermal fluctuations present within the domains, the systems evolve into stripe-like configurations at late times \cite{basu2017phase}.
\begin{figure}[t!]
\centering
\includegraphics*[width=0.45 \textwidth]{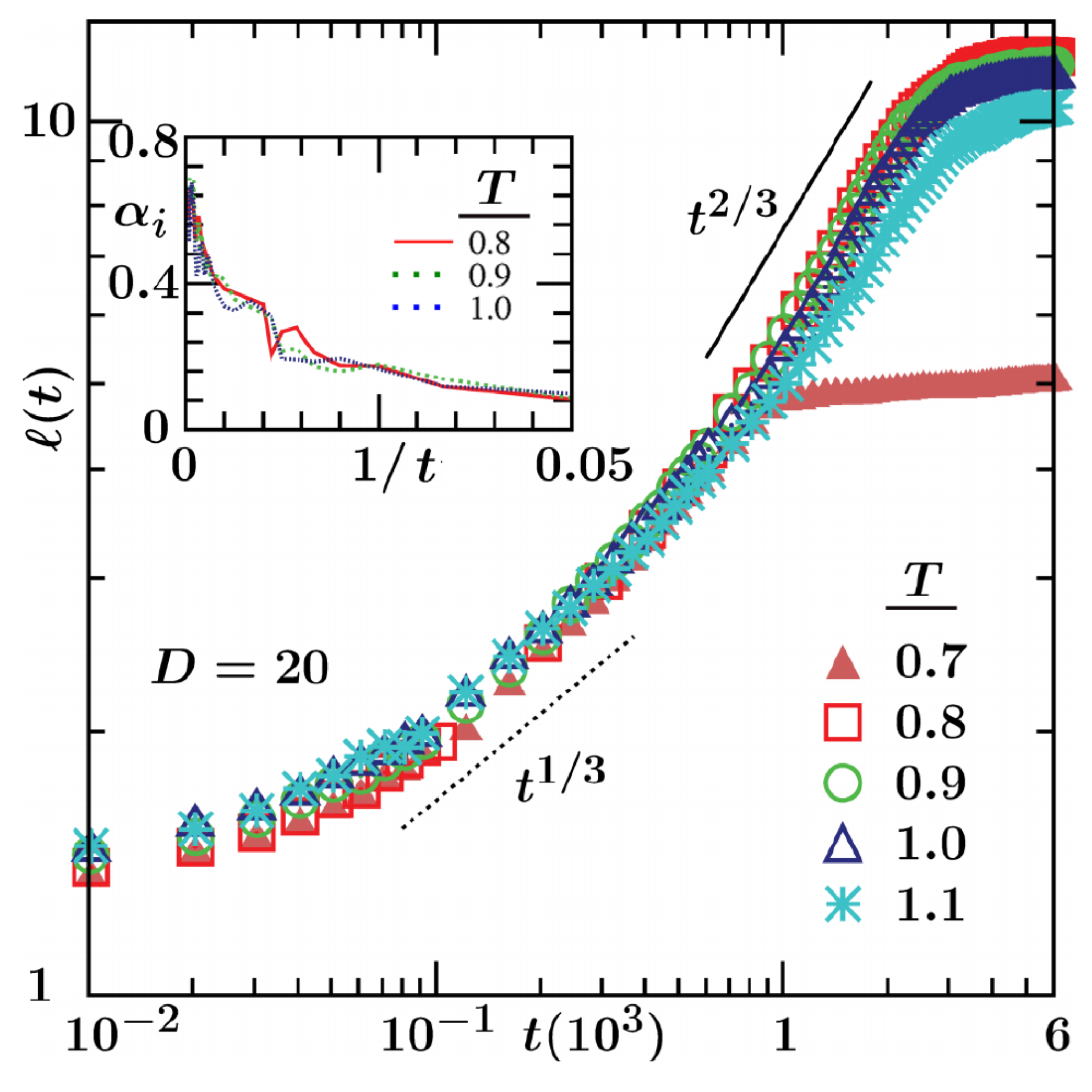}
\caption{\label{fig5} Plots of the characteristic length scale $\ell(t)$ ($D=20, L=200$), versus time, on a log-log scale, for different $T$. The dotted line corresponds to diffusive ($\alpha=1/3$) and solid line represents inertial hydrodynamic ($\alpha=2/3$) growth laws. Inset: $\alpha_{i}$, obtained from the data in the main frame for three different temperatures, is plotted against $1/t$.
}
\end{figure}

\par
Qualitative results on the growth of these patterns at different final temperatures ($0.7 \leq T \leq 1.1$) are shown in Fig.~\ref{fig5}, where $\ell(t)$ is plotted as a function of $t$ on a double-log scale. The first notable observation is that $\ell(t)$ for the lowest considered temperature ($T = 0.7$) saturates at a significantly smaller value compared to the other temperatures, consistent with the snapshot in Fig.~\ref{fig4}. For $T = 0.7$, after an initial diffusive growth regime ($\alpha = 1/3$, indicated by the dotted black line) at early times, domain sizes show only moderate increase with time, and the saturated domain length is much smaller than for the higher temperatures. For the other temperatures ($T \geq 0.8$), at late times, the inertial hydrodynamic regime ($\alpha = 2/3$) becomes evident, as indicated by the agreement of the data with the solid black line in Fig.~\ref{fig5}. This is further confirmed by plotting the instantaneous exponent $\alpha_i$ versus $1/t$ in the inset of Fig.~\ref{fig5}, which shows that $\alpha_i$ approaches $\simeq 2/3$ in the asymptotic limit. The domain growth eventually saturates at nearly the same timescale for all cases. As seen in Fig.~\ref{fig5}, the saturated domain length, $\ell_{\rm sat}$, for $T \geq 1.0$ is slightly smaller than that for $T = 0.8$ and $0.9$, which can be attributed to stronger thermal fluctuations. One could attempt to estimate $\ell(t)$ by filtering out noise, as done for solid binary mixtures in~\cite{majumder2010,majumder2011diffusive}. However, such an approach must be applied cautiously because at higher temperatures the noise clusters, typically of the size of the equilibrium correlation length $\xi$, are not negligible compared to the growing domain size $\ell(t)$, and neglecting them may introduce unwanted artifacts. Therefore, we refrain from performing this exercise.
\begin{figure}[t!]
\centering
~\includegraphics*[width=0.46\textwidth]{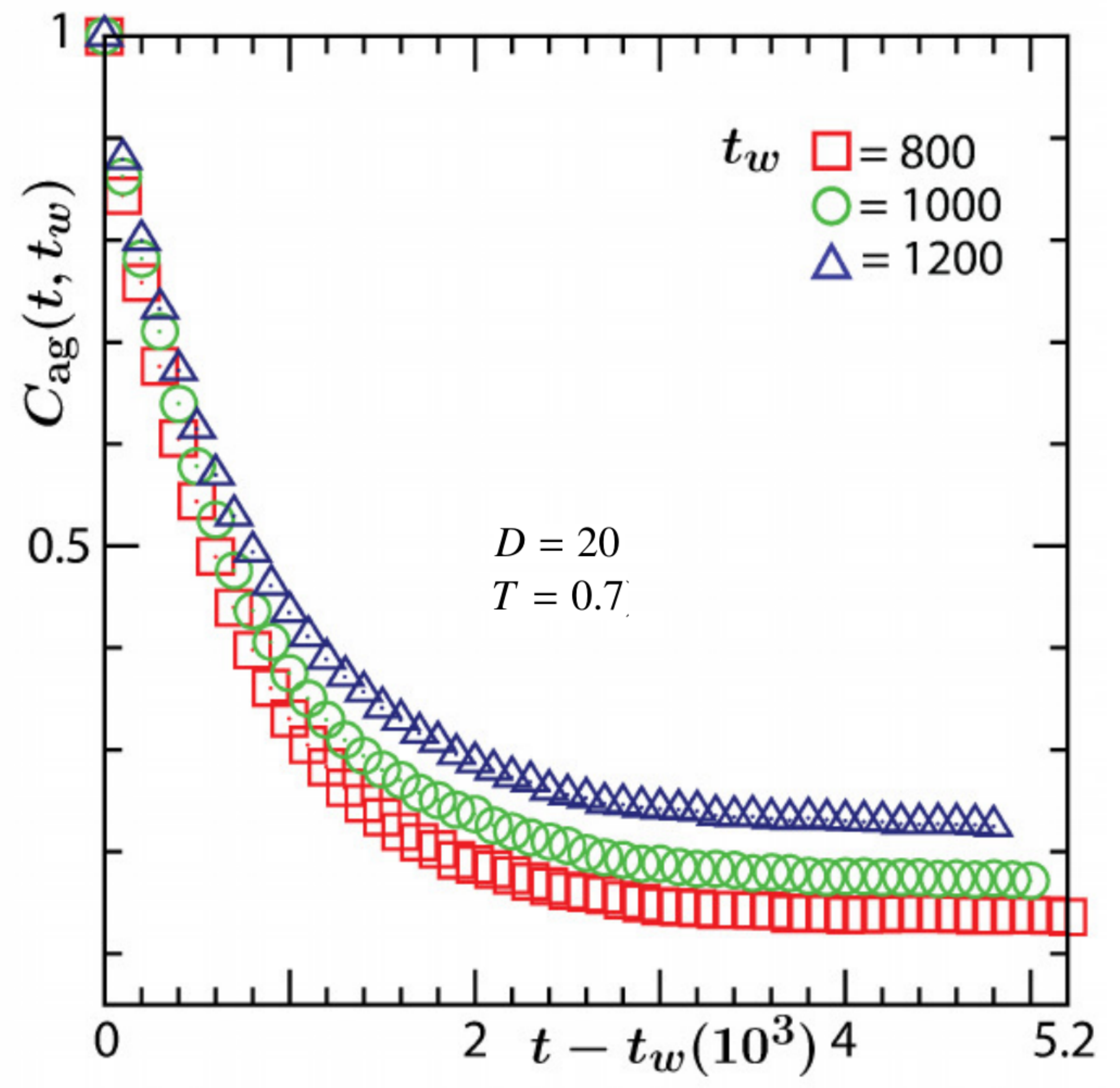}
\caption{\label{fig6} Demonstration of the breaking of the time translational invariance in autocorrelation functions $C_{\rm{ag}}(t,t_w)$, calculated at different waiting times $t_w$ (for  $D=20$ and $L=200$, at $T=0.7$), when plotted versus the translated time, $t-t_w$.} 
\end{figure}

\subsection{Aging and Dynamical Scaling}\label{aging}
In the previous subsection, we examined the kinetics using a single-time quantity, namely the average domain size $\ell(t)$. As noted in Sec.~\ref{intro}, another important aspect of phase transition kinetics involves multi-time quantities, which provide insights into aging and associated scaling behavior \cite{fisher1988}. In this subsection, we investigate this aspect through $C_{\rm ag}(t,t_w)$, defined earlier in Eq.~\eqref{autocor}. Unless stated otherwise, all results presented here correspond to $T = 0.8$.

\par
In Fig.~\ref{fig6} we plot $C_{\rm{ag}}(t,t_{w})$, for a few different $t_w$, versus the translated time $t-t_{w}$, with $D=20, L=200$. The data clearly demonstrate the absence of time translation invariance, indicating the existence of aging \cite{fisher1988}. This can be further confirmed by extracting a relaxation time $\tau$ for different $t_w$. Roughly, this exercise yields a behavior $\tau \sim t_w$ which not only confirms slow relaxation but also indicates the presence of simple aging rather than 
any special aging ~\cite{henkelbook}. Note that here and in all the subsequent exercises, the $t_w$ values are always chosen to be in the range where the domain growth follows $\ell(t)\sim t^{2/3}$ behavior.

\par
\begin{figure}[t!]
\centering
\includegraphics*[width=0.45 \textwidth]{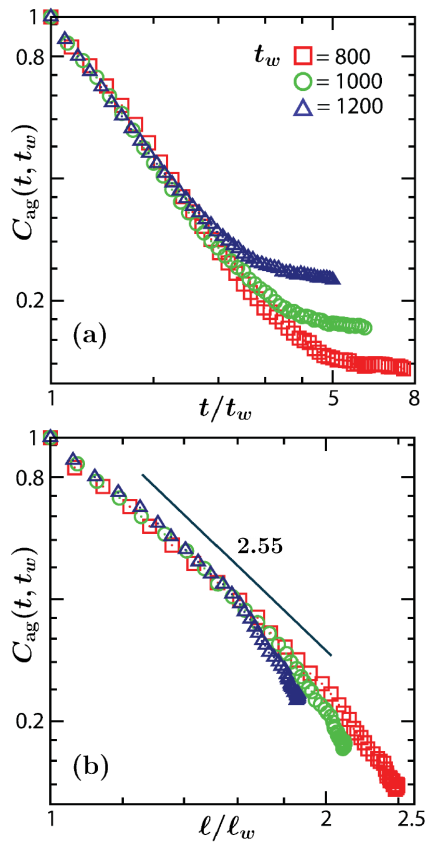}
\caption{\label{fig7} Plots of the same autocorrelation functions $C_{\rm{ag}}(t,t_w)$, shown in Fig.~\ref{fig6}, versus (a) the rescaled time $t/t_w$ and (b) $x$ ($=\ell/\ell_{w}$), on a double-log scale. The solid line denotes a power-law decay with an exponent consistent with $\lambda=2.55$. The deviations of the data sets at larger abscissa values signify the onset of finite-size effects.
}
\end{figure}

The next important aspect to examine is the dynamical scaling described in Eq.~\eqref{corscaling}. In Fig.~\ref{fig7}(a), we plot $C_{\rm ag}(t,t_w)$ as a function of $t/t_w$, and in Fig.~\ref{fig7}(b), as a function of $x = \ell(t)/\ell(t_w)$, both on a double-logarithmic scale. In both representations, the data from different $t_w$ collapse nicely. We note, however, that poor data quality when plotting against $t/t_w$, particularly in soft matter systems~\cite{Stanley2002}, has sometimes led to claims of sub-aging or super-aging, where $t/t_w$ is replaced by $t/t_w^\mu$ with $\mu < 1$ or $\mu > 1$. It has been argued~\cite{Park2010,christiansen2017} that plotting the data with respect to $x = \ell(t)/\ell(t_w)$ often provides a better collapse and supports a simple aging scenario. Moreover, recent studies have emphasized the need for caution in interpreting apparent sub-aging behavior~\cite{christiansen2020aging,christiansen2025finite}.
\begin{figure}[t!]
\centering
\includegraphics*[width=0.45 \textwidth]{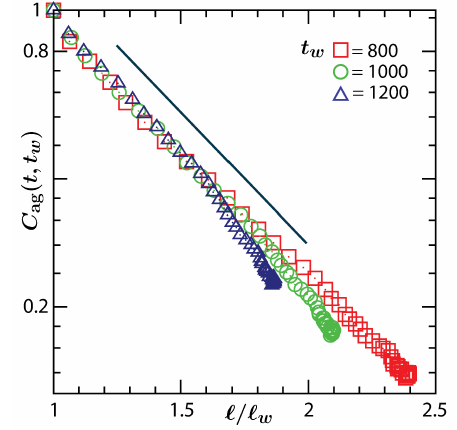}
\caption{\label{fig8} Previously displayed autocorrelation functions, $C_{\rm{ag}}(t,t_w)$, are shown against  $x$ ($=\ell/\ell_{w}$), for different waiting times ($t_w$) on a semi-log scale to check the validity of an exponential decay. The continuous line provides a guide for an exponentially decaying function.}
\end{figure}

\par
Next, we examine the form of the scaling behavior shown in Fig.~\ref{fig7}(b). In this double-log plot, the data do not appear strictly linear, as would be expected for a pure power-law decay; instead, the slope changes gradually with $x$. Nevertheless, a power-law-like trend becomes more apparent at larger values of $x$, suggesting the presence of finite-time corrections to an underlying power-law scaling \cite{Midya2014}. At the same time, it is also worthwhile to consider whether the scaling could follow an exponential decay, as discussed in Ref.~\cite{roy2019aging} and references therein. In Fig.~\ref{fig8}, we plot $C_{\rm ag}(t,t_w)$ versus $x$ on a semi-log scale, where the data appear closer to linear. However, caution is needed before concluding an exponential form, as demonstrated in the study of aging in bulk fluids~\cite{roy2019aging}, because the range of the abscissa is quite limited. Achieving a broader range is inherently difficult due to the system’s characteristics and dynamics. In Ref.~\cite{roy2019aging}, the apparent exponential behavior was ruled out through advanced finite-size scaling (FSS) analysis~\cite{Midya2014,Midya2015,majumder2016polymer,majumder2017kinetics,vadakkayil2019,roy2019aging}, which not only hinted towards the power-law scaling but also provided a reliable estimate of the decay exponent $\lambda$. Following this approach, we pursue a similar analysis here.   

\par 
To proceed, one must begin with a functional form of the autocorrelation function $C_{\rm ag}(t,t_w)$, which can be confirmed through finite-size scaling (FSS). Relying on the previous studies~\cite{Midya2015,roy2019aging}, we write:
\begin{eqnarray}\label{empscaling}
C_{\rm{ag}}(t,t_{w}) =Bx^{-\lambda}\exp(-A_{c} /x),
\end{eqnarray}
where $A_c$ and $B$ are constants. Note that in Eq.~\eqref{empscaling} the exponential factor, which takes care of the early-time correction, was constructed empirically in  Ref.~\cite{Midya2014} while studying aging during ferromagnetic ordering. This function converges to power-law scaling ($\sim x^{-\lambda}$), which was originally proposed in Ref.~\cite{fisher1988}, in the asymptotic limit of $x \to \infty$. The FSS function, $Y$, with the form of $C_{\rm{ag}}(t,t_w)$ in Eq.~\eqref{empscaling}, reads \cite{vadakkayil2019}
\begin{eqnarray}\label{FSSscaling}
Y=C_{\rm{ag}}(t,t_{w})\exp\left(\frac{By}{y_{w}}\right)y^{\lambda}_{w},
\end{eqnarray}
where $y=L/\ell$ and $y_{w}=L/\ell_{w}$. The changeover from $x$ to $y$ has been motivated by the following fact. Analogous to the dimensionless quantity $L/\xi$ in critical phenomena ~\cite{Fisherbook,Privmanbook}, where $\xi$ is the equilibrium correlation length, $L/\ell$ can serve the purpose of an appropriate FSS variable \cite{Midya2014}. $Y$ is expected to be independent of system size as $y$ is a dimensionless variable. One can, therefore, expect a collapse of data in a $Y$ vs $y$ plot, from various system sizes. Now, in the absence of any finite-size effects, i.e., in the limit $y \rightarrow \infty$ ($L \gg \ell$), Eq.\ \eqref{FSSscaling} will reduce to Eq.\ \eqref{empscaling} provided \cite{vadakkayil2019}
\begin{eqnarray}\label{FSSscaling1}
Y \sim y^{\lambda}.
\end{eqnarray}

\par 
In the context of aging phenomena, an important aspect of the FSS analysis is that one need not simulate systems of different sizes \cite{Midya2015}. In Fig.~\ref{fig7}(b), which shows the autocorrelation data for fixed $D$ and $L$, finite-size effects become noticeable at higher $t_w$ for smaller values of $x$. This behavior is analogous to the faster appearance of finite-size effects in physically smaller systems for a given $t_w$. Therefore, $t_w$ effectively plays the role ~\cite{vadakkayil2019} of $L$, allowing us to avoid simulations for multiple system sizes, which would be computationally demanding. For our FSS analysis, we use data from different $t_w$ values for fixed system dimensions with $D = 20$ and $L = 200$.
\begin{figure}[t!]
\centering
\includegraphics*[width=0.45 \textwidth]{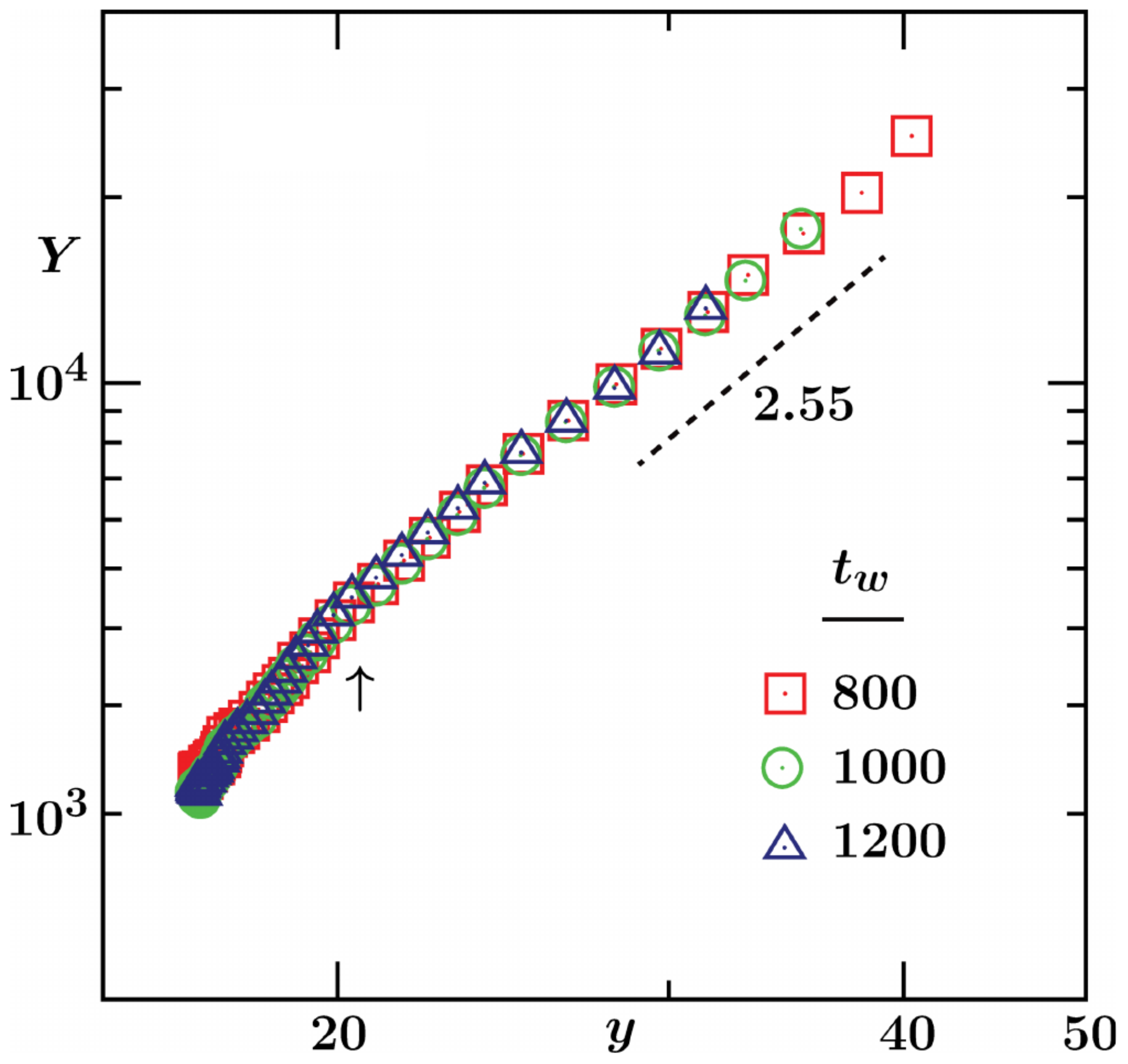}
	\caption{\label{fig9} Illustration of the finite-size scaling exercise for $D=20$ and $T=0.8$, using the $C_{ag}(t,t_w)$ data for different $t_w$, as indicated. The dashed line represents a power-law, shown in Eq.\ \eqref{FSSscaling1}, having an exponent $\lambda = 2.55$. The arrow approximately marks the onset of the finite-size effect.}
\end{figure}

\par 
During the FSS analysis, one tunes the value of $\lambda$ (as well as $A_c$, noting that $B$ is kept fixed and is not an adjustable parameter) to achieve the optimal collapse of the data. For the system parameters considered here, the best collapse is obtained for $\lambda = 2.55$. In Fig.~\ref{fig9}, we show the corresponding FSS plot using data for three different $t_w$ values. At larger values of $y$, the master curve agrees well with the form given in Eq.~\eqref{FSSscaling1}. The deviation from the power-law scaling is indicated by an arrow, roughly corresponding to the point where finite-size effects start to appear. This estimate of the finite-size crossover is consistent with that inferred from the growth data. The persistence of the power-law behavior of $Y$ up to the onset of finite-size effects reinforces the correctness of the exponential correction in the power-law decay of $C_{\rm ag}(t,t_w)$, as noted in Eq.~\eqref{empscaling}.

\par
The value of the exponent $\lambda=2.55$ is in accordance with the lower bound obtained by Fisher and Huse \cite{fisher1988}. Recall that in the present case, the system has quasi-one-dimensional geometry, i.e., we have effectively $d=1$. The stricter lower bound, which was proposed by YRD, suggests that $\lambda > 3/2$, using the fact that for $d=1$ one expects \cite{yeung1988scaling,Furukawa1989k} $\beta\simeq 2$. Our previous study ~\cite{basu2017phase} confirmed that in the present quasi-one-dimensional case, indeed $\beta=2$, which reaffirms that the YRD bound in the present case is in fact $3/2$. The value $\lambda=2.55$ certainly respects that latter bound.
\begin{figure}[t!]
\centering
\includegraphics*[width=0.45 \textwidth]{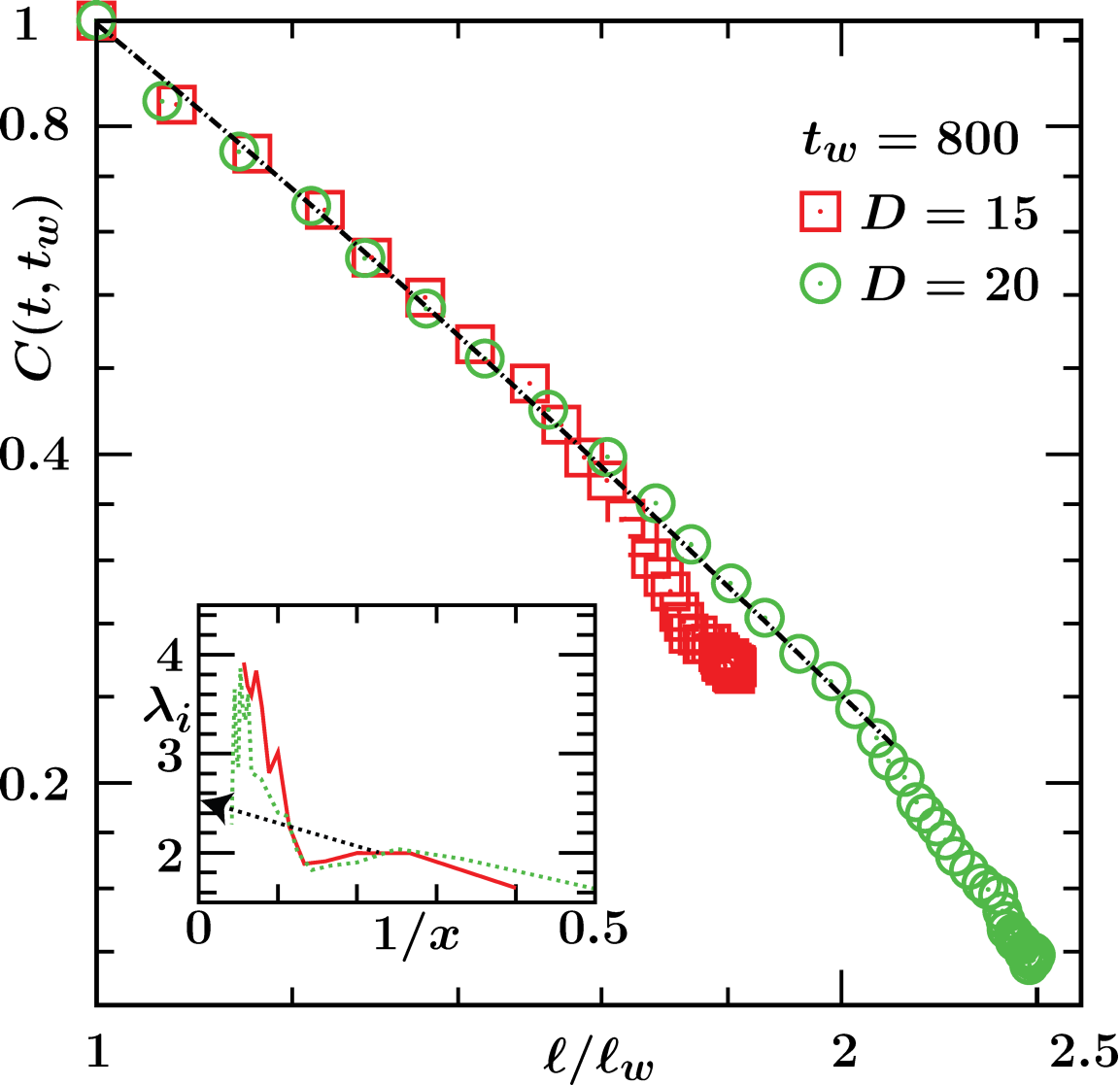}
\caption{\label{fig10} Plots of $C_{\rm{ag}}(t,t_{w})$, versus $x$, with fixed $t_{w}=800$, for two different $D$ as mentioned, with $L=200$. The dashed-dotted line corresponds to Eq.\ \eqref{empscaling} with $\lambda=2.55$. The inset shows corresponding plots of the instantaneous exponent $\lambda_i$, calculated via Eq.\ \eqref{instexp2}, as a function of $1/x$.  The dotted arrow-headed line provides a possible guide about how $\lambda_i$ may approach the asymptotic value $\lambda=2.55$.}
\end{figure}

\par
To check the effect of cylinder diameter on the validity of Eq.~\eqref{empscaling} with $\lambda = 2.55$, we plot $C_{\rm ag}(t,t_w)$ for two values of $D$ in the main panel of Fig.~\ref{fig10}, with $t_w = 800$ fixed. The data for both diameters coincide until the curve for $D = 15$ deviates due to the earlier onset of finite-size effects, i.e., freezing into the stripe state [see Fig.~\ref{fig3}]. The dashed-dotted line in Fig.~\ref{fig10} represents a fit to Eq.~\eqref{empscaling}, using $\lambda = 2.55$ and excluding the region affected by finite-size effects. There is good agreement with the simulation data for both $D$ until finite-size effects set in, confirming the accuracy of $\lambda = 2.55$, which is expected to hold as long as growth remains anisotropic. This asymptotic value can be further verified by calculating the instantaneous exponent~\cite{vadakkayil2019}
\begin{eqnarray}\label{instexp2}
\lambda_{i}=-\frac{ d \ln [C_{\rm{ag}}(t,t_{w})]}{d \ln x}.
\end{eqnarray}
Extrapolations of the plots of $\lambda_i$ as a function of $1/x$ [$\ell(t)/\ell(t_w)$] to the limit $1/x \to 0$ are expected to provide the asymptotic value. However, in our case, this procedure does not yield a conclusive result, as shown in the inset of Fig.~\ref{fig10}. Obtaining well-behaved data for $\lambda_i$ would require significantly greater computational resources. The data in the inset exhibit considerable statistical fluctuations, and therefore, we refrain from drawing a firm conclusion. Nevertheless, the main panel already provides a strong indication that the finite-size unaffected data are in satisfactory agreement with the value $\lambda = 2.55$.

\begin{figure}[t!]
\centering
\includegraphics*[width=0.45 \textwidth]{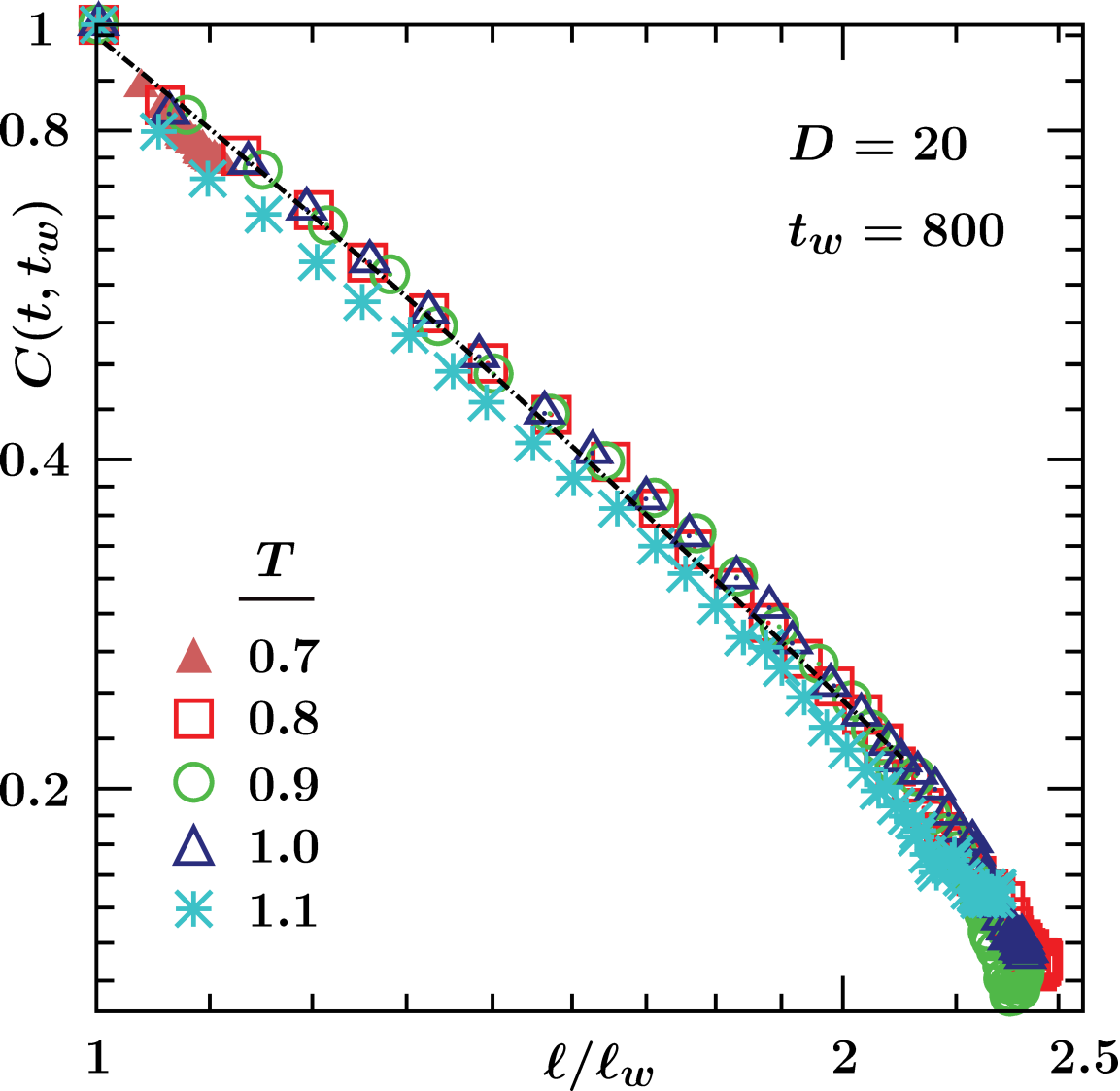}
\caption{\label{fig11} Plots of $C_{\rm{ag}}(t,t_{w})$, versus $x$, from five different temperatures $T$, as mentioned, with $D=20$, $L=200$ and fixed $t_w=800$. The dashed-dotted line corresponds to Eq.\ \eqref{empscaling} with $\lambda=2.55$. }
\end{figure}
\par 
Finally, we investigate the behavior of $C_{\rm ag}(t,t_w)$ for different quench temperatures. In Fig.~\ref{fig11}, plotted using a log-log scale, we show $C_{\rm ag}(t,t_w)$ for various $T$ with $D = 20$, $L=200$ and $t_w = 800$. All data sets for the different final temperatures collapse reasonably well onto each other, indicating a temperature-independent behavior. The dashed-dotted line represents Eq.~\eqref{empscaling} with $\lambda = 2.55$. The satisfactory collapse of the data across different $T$ suggests the presence of a universal form. In this context, a scaling analysis demonstrating a universal finite-size scaling function, similar to that observed for domain growth in phase separation of multicomponent mixtures~\cite{majumder2018universal}, could be explored.

\section{Summary and discussions}\label{conclusion}
Using MD simulations, we have obtained results for domain growth and aging phenomena during phase separation in binary (A+B) liquid mixtures confined within cylindrical nanopores. Regardless of the quench temperature, the relaxation is characterized by the formation of stripe patterns, where domains of A and B species alternate along the cylinder axis. The growth of these stripes follows a scaling behavior which, unlike in bulk fluids, exhibits only two distinct regimes. After an initial diffusive growth \cite{lifshitz1961kinetics} with $\alpha = 1/3$, a faster hydrodynamic growth emerges, for which the estimated exponent $\alpha = 2/3$ is consistent with the inertial hydrodynamic growth \cite{bray2002theory} observed in three-dimensional bulk fluids. This crossover from diffusive to inertial hydrodynamic growth appears to be temperature independent. This behavior differs from that in three-dimensional bulk fluids, where an intermediate viscous hydrodynamic regime is typically observed~\cite{bray2002theory,Ahmad2010,majumder2011VL} with a power-law exponent $\alpha = 1$.

\par
In this paper, our primary focus was to investigate the aging dynamics \cite{fisher1988} associated with the nonequilibrium process of phase segregation. We employed the autocorrelation function, $C_{\rm ag}(t,t_w)$, as a probe. Analysis of our results, using an advanced finite-size scaling approach \cite{vadakkayil2019}, provides evidence for simple aging behavior, with $C_{\rm ag}(t,t_w)$ exhibiting a power-law scaling with respect to $x = \ell(t)/\ell(t_w)$, the ratio of domain sizes at times $t$ and $t_w$, respectively. This power-law behavior [$C_{\rm ag}(t,t_w) \sim x^{-\lambda}$] is consistent with similar observations in bulk fluids~\cite{roy2019aging}, although the autocorrelation exponent $\lambda$ differs. We further confirmed that the observed value $\lambda = 2.55$ is independent of the quench temperature. Hence, we conclude that while the relaxation dynamics inside a nanopore are significantly altered—the power-law scaling of $C_{\rm ag}(t,t_w)$ remains robust.
 
\par
Recent studies across different systems have revealed a global trend -- when the spatial dimension remains fixed, the autocorrelation exponent tends to be smaller for systems exhibiting faster growth~\cite{ghosh2024nonuniversal}. In contrast, here the value $\lambda \simeq 2.55$ is smaller than $\lambda \simeq 4$ observed for viscous hydrodynamic growth with $\alpha = 1$ in $d = 3$. Given that the cylinder diameter is relatively large, one might argue that the system is closer to three-dimensional than one-dimensional. However, structural characteristics and the behavior at both short and long wave vectors regimes strongly indicate a one-dimensional nature. Therefore, the smaller value of $\lambda$ observed here, despite a lower $\alpha$, should be attributed to the difference in dimensionality.

\par
The results presented here arise from a study of systems within cylindrical geometry, which indeed imparts a quasi-one-dimensional character to the system. From this viewpoint, it would be interesting to examine analogous geometrical restrictions that produce a quasi-two-dimensional system, such as thin film geometries~\cite{das2006molecular}. Similar to the present case, growth kinetics in such systems also lack a viscous hydrodynamic regime. However, there are currently no studies investigating the corresponding aging behavior, which we plan to explore in future work.

\section{Acknowledgment} 
S.M.\ acknowledges the Anusandhan National Research Foundation (ANRF), Govt.\ of India through a Ramanujan Fellowship (File No.\ RJF/2021/000044). S.K.D.\ acknowledges partial supports from Department of Science and Technology, India; Marie-Curie Actions Plan of European Commission (FP7-PEOPLE-2013-IRSES Grant No.\ 612707, DIONICOS); and International Centre for Theoretical Physics, Italy. He also thanks S. Chandel for technical help. S.B.\ and R.P.\ thanks IACS, Kolkata, for financial support. 
%
\end{document}